\documentclass[aps,prl,twocolumn,showpacs,superscriptaddress,citeautoscript,floatfix,longbibliography]{revtex4-1}
\usepackage{amsfonts}
\usepackage{amsmath}
\usepackage{amssymb}
\usepackage{graphicx}
\usepackage{dcolumn}
\usepackage{bm}\let\vec\bm
\usepackage[svgnames]{xcolor}
\usepackage[
	colorlinks=True,linkcolor=DarkRed,citecolor=ForestGreen,urlcolor=MediumBlue,
	pdfstartview=FitH,bookmarks=False,pdfpagemode=UseNone
]{hyperref}

\begin{document}

\title{\boldmath Tilted vortex cores and superconducting gap anisotropy in 2H-NbSe$_2$}

\author{J.A. Galvis} 
\affiliation{Laboratorio de Bajas Temperaturas y Altos Campos Magn\'eticos, Unidad Asociada UAM/CISC, Departamento de F\'isica de la Materia Condensada, Instituto de Ciencia de Materiales Nicol\'as Cabrera, Center for Condensed Matter Physics (IFIMAC), Universidad Aut\'onoma de Madrid, E-28049 Madrid, Spain}
\affiliation{Departamento de ciencias naturales, Facultad de ingenieria y ciencias b{\'a}sicas, Universidad Central, Bogot\'a 110311, Colombia}
\author{E. Herrera}
\affiliation{Laboratorio de Bajas Temperaturas y Altos Campos Magn\'eticos, Unidad Asociada UAM/CISC, Departamento de F\'isica de la Materia Condensada, Instituto de Ciencia de Materiales Nicol\'as Cabrera, Center for Condensed Matter Physics (IFIMAC), Universidad Aut\'onoma de Madrid, E-28049 Madrid, Spain}
\affiliation{Departamento de F\'isica, Universidad Nacional de Colombia, Bogot\'a 111321, Colombia}
\author{C. Berthod}
\affiliation{Department of Quantum Matter Physics, University of Geneva, 24 quai Ernest-Ansermet, 1211 Geneva, Switzerland}
\author{S. Vieira}
\affiliation{Laboratorio de Bajas Temperaturas y Altos Campos Magn\'eticos, Unidad Asociada UAM/CISC, Departamento de F\'isica de la Materia Condensada, Instituto de Ciencia de Materiales Nicol\'as Cabrera, Center for Condensed Matter Physics (IFIMAC), Universidad Aut\'onoma de Madrid, E-28049 Madrid, Spain}
\author{I. Guillam\'on}
\affiliation{Laboratorio de Bajas Temperaturas y Altos Campos Magn\'eticos, Unidad Asociada UAM/CISC, Departamento de F\'isica de la Materia Condensada, Instituto de Ciencia de Materiales Nicol\'as Cabrera, Center for Condensed Matter Physics (IFIMAC), Universidad Aut\'onoma de Madrid, E-28049 Madrid, Spain}
\author{H. Suderow}
\affiliation{Laboratorio de Bajas Temperaturas y Altos Campos Magn\'eticos, Unidad Asociada UAM/CISC, Departamento de F\'isica de la Materia Condensada, Instituto de Ciencia de Materiales Nicol\'as Cabrera, Center for Condensed Matter Physics (IFIMAC), Universidad Aut\'onoma de Madrid, E-28049 Madrid, Spain}

\date{November 22, 2017}

\begin{abstract}
Superconducting vortex cores have been extensively studied for magnetic fields applied perpendicular to the surface by mapping the density of states (DOS) through Scanning Tunneling Microscopy (STM). Vortex core shapes are often linked to the superconducting gap anisotropy---quasiparticle states inside vortex cores extend along directions where the superconducting gap is smallest. The superconductor 2H-NbSe$_2$ crystallizes in a hexagonal structure and vortices give DOS maps with a sixfold star shape for magnetic fields perpendicular to the surface and the hexagonal plane. This has been associated to a hexagonal gap anisotropy located on quasi two-dimensional Fermi surface tubes oriented along the $c$ axis. The gap anisotropy in another, three-dimensional, pocket is unknown. However, the latter dominates the STM tunneling conductance. Here we measure DOS in magnetic fields parallel to the surface and perpendicular to the $c$ axis. We find patterns of stripes due to in-plane vortex cores running nearly parallel to the surface. The patterns change with the in-plane direction of the magnetic field, suggesting that the sixfold gap anisotropy is present over the whole Fermi surface. Due to a slight misalignment between the vector of the magnetic field and the surface, our images also show outgoing vortices. Their shape is successfully compared to detailed calculations of vortex cores in tilted fields. Their features merge with the patterns due to in plane vortices, suggesting that they exit at an angle with the surface. Measuring the DOS of vortex cores in highly tilted magnetic fields with STM can thus be used to study the superconducting gap structure.
\end{abstract}
\maketitle 

The superconducting compound 2H-NbSe$_2$ is considered as a prototypical example of a material with highly anisotropic superconducting properties---for instance, the upper critical field is three times larger when the field is applied in-plane than when it is perpendicular to the planes of the hexagonal crystalline structure \cite{Corcoran94}. The Fermi surface of 2H-NbSe$_2$ has two tubes around the $\Gamma$ and K points due to bands derived from Nb orbitals which are nearly two-dimensional and a small three-dimensional pocket centered on the $\Gamma$ point due to bands derived from Se orbitals \cite{Corcoran94,Janssen98,Kiss07,Rahn12,Johannes06}. A charge density wave (CDW) opens below 30~K in 2H-NbSe$_2$, coexisting with superconductivity below $T_c=7.2$\,K \cite{Rahn12}. STM measurements in magnetic fields perpendicular to the hexagonal planes provide DOS maps giving sixfold vortex cores and the atomic scale tunneling conductance has a sixfold modulation \cite{Caroli64,Hess89,Hess90,Hayashi98,Guillamon08PRB}. Photoemission experiments suggest that the CDW induces the sixfold anisotropic superconducting gap \cite{Kiss07,Rahn12}. This is reinforced by STM measurements on the superconducting gap in the isostructural compound 2H-NbS$_2$, which has a similar $T_c$ and no CDW and is isotropic in-plane \cite{NbS2}.

Photoemission experiments also suggest that the superconducting gap anisotropy is located in the Nb tubes and do not address the three dimensional Se derived band \cite{Kiss07,Rahn12}. On the other hand, penetration depth studies show that the gap at the Se sheet is large \cite{Fletcher07}. Furthermore, the Se sheet plays an essential role in most tunneling experiments \cite{Johannes06}. Tunneling occurs preferentially through the last layer that consists of the hexagonal Se atom lattice \cite{PhysRevB.92.134510,Johannes06}. It is however yet unclear what is the contribution of the electronic properties derived from Se to the superconducting gap anisotropy. To study this issue, measurements of the vortex core shape on a surface perpendicular to the $c$ axis would be useful. However, no vortex imaging can then be made, because the surface is highly irregular due to the sheet-like structure of samples of 2H-NbSe$_2$. Here we study the usual surface of 2H-NbSe$_2$, parallel to the hexagonal plane, and apply the magnetic field perpendicular to the $c$ axis, nearly parallel to the surface. We visualize the DOS patterns produced on the surface by in-plane vortex cores and relate these to the out of plane gap anisotropy. We also calculate the spatial dependence of the DOS in tilted vortex cores.

Hess \textit{et al.}\ measured the vortex lattice of 2H-NbSe$_2$ with STM in tilted magnetic fields \cite{Hess92,Hess94}. They focused on the structure of the vortex lattice and found a distortion of the hexagonal lattice compatible with the anisotropy of the upper critical field, as well as a rotation of the orientation of the vortex lattice that is consistent with the anisotropic London theory \cite{Campbell88,Gammel94,Kogan95}. When the magnetic field was close to being parallel to the hexagonal planes, they observed elongated vortex cores, instead of the sixfold cores seen in perpendicular fields, and a peculiar pattern of stripes. More recently, this pattern of stripes was addressed in STM measurements with the field exactly in-plane focusing on the influence of the structure of the vortex lattice and of screening currents \cite{Fridman11,Fridman13}. Until now, there is no systematic study of the vortex core shape with the in-plane direction of the magnetic field and the origin of the observed stripe pattern remains unclear.

We perform a detailed study as a function of the direction and bias voltage and show that the pattern of stripes is due to bound states of subsurface vortices. Furthermore, we find that the pattern changes between a two-fold vs three-fold structure depending on the direction of the in-plane magnetic field with respect to the crystalline direction in the hexagonal plane. The images strongly change with the bias voltage, with the stripes disappearing when reaching the gap edge. We perform microscopic calculations of the DOS maps and their bias voltage dependence and find that the shape of in-plane vortex cores depends on the direction of the magnetic field with respect to the in-plane crystal lattice direction.

\section*{Results}

\begin{figure}[tb]
\centering
\includegraphics[width=\columnwidth]{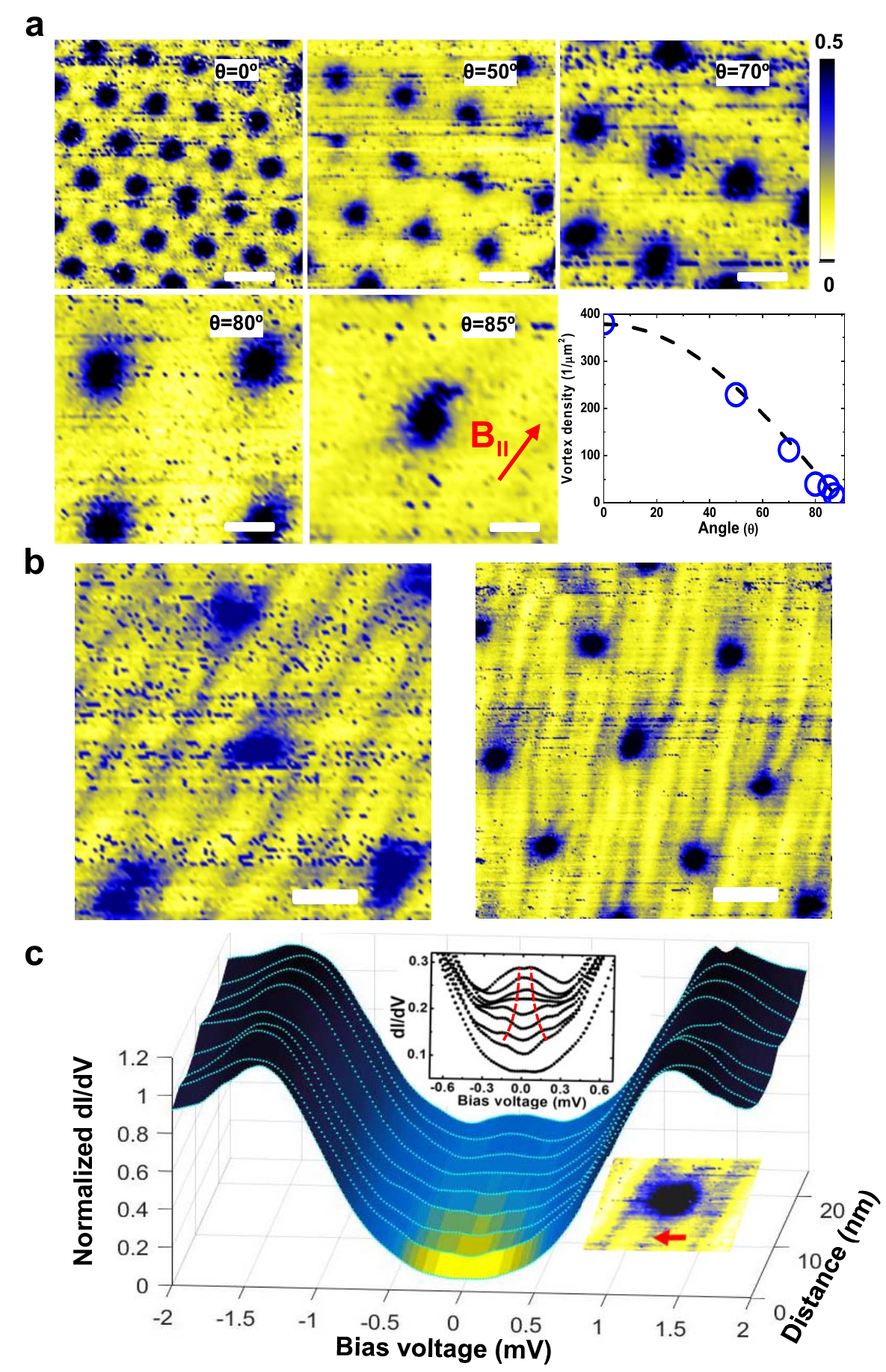}
\caption{\textbf{Tilted vortex cores}. (\textbf{a}), The vortex lattice as a function of the polar angle ($\theta$) at $0.6$ T and fixed azimuthal ($\varphi=0^{\circ}$) angle. Images show the zero-bias normalized conductance (normalized conductance units, NCU). Scale bar is of 50 nm. Note the increase in the vortex core size when tilting the magnetic field. In the lower right panel we show the density of the vortex lattice at the surface (points) compared to expectations using the anisotropic London theory \cite{Campbell88}. (\textbf{b}), Zero bias conductance maps with polar angle close to $\theta = 90^{\circ}$ (see text and methods section for the uncertainty in $\theta$) for two different azimuthal angles $\varphi=0^{\circ}$ (left panel) and $\varphi=30^{\circ}$ (right panel). Color scale of both figures is given in top right corner of \textbf{a}. (\textbf{c}), Normalized tunneling conductance curves when crossing a stripe along the red line shown in the lower right inset. Upper inset shows the evolution of the conductance close to zero bias, with a dashed red line marking the splitting of the zero-bias peak. All data are taken at $T=150$~mK.}
\label{f1}
\end{figure}

In Fig.~\ref{f1}a we show the vortex lattice at $0.6~\mathrm{T}$ with varying polar angle $\theta$. The magnetic field is tilted at a fixed azimuthal angle $\varphi = 0^{\circ}$, along an in-plane crystalline axis. That is, along the nearest neighbor axis in the hexagonal plane of the atomic surface plane (given by Se atoms, see supplementary information and Refs.~\cite{Guillamon08PRB,Hess90}). For all tilts we observe a peak in the tunneling conductance due to Caroli--de Gennes--Matricon bound states at the center of the vortex cores, similar to the one observed at perpendicular magnetic fields \cite{Caroli64,Hess89,Guillamon08PRB,Hess90}. However, there are remarkable differences in tilted fields.

The vortex core as observed in the zero bias tunneling conductance maps continuously increases its size when tilting the magnetic field and acquires a two-fold shape at nearly in-plane magnetic fields. For these high tilt angles, we observe a pattern of stripes in the zero bias tunneling conductance maps (Fig.~\ref{f1}b).

In Fig.~\ref{f1}c we show the bias-voltage dependence of the tunneling conductance when crossing a stripe. We start from a gapped behavior in-between vortices and observe a small zero-bias peak when we cross the stripe. The zero-bias peak splits into two peaks at nonzero bias when leaving the center of the stripe (Fig.~\ref{f1}c inset, see also Fig.~\ref{fs5} of the supplementary information), as expected for Caroli--de Gennes--Matricon bound states \cite{Caroli64,Hess89,Guillamon08PRB,Hess90}. Thus, the stripes result from vortex-core bound states from subsurface vortices oriented along the tilt direction. Accordingly, the stripes always follow the direction of the magnetic field.

In addition, the pattern of stripes changes with the in-plane azimuthal angle $\varphi$ with respect to the in-plane hexagonal crystal lattice. For the field along a crystal axis, $\varphi=0^{\circ}$ (left panel of Fig.~\ref{f1}b), the amount of stripes is about half the amount of stripes observed for $\varphi=30^{\circ}$ (right panel of Fig.~\ref{f1}b, this is at $90^{\circ}$ (or equivalently at $30^{\circ}$) from the nearest neighbor Se--Se direction of the surface atomic lattice, see supplementary information and Refs.~\cite{Guillamon08PRB,Hess90}). The distance between vortices in the bulk is about 90~nm for a parallel magnetic field of 0.6~T. In Fig.~\ref{f1}b (left panel), when the magnetic field is along a crystalline axis ($\varphi = 0^{\circ}$), we obtain 90~nm for the average distance between stripes. On the other hand, when the tilt of the magnetic field is in between crystalline axis ($\varphi = 30^{\circ}$), we observe an average distance between stripes of 40~nm, about half the value found for $\varphi = 0^{\circ}$ (Fig.~\ref{f1}b, right panel). This suggests that there is one stripe per in-plane vortex for $\varphi = 0^{\circ}$, but two stripes for $\varphi = 30^{\circ}$ per in-plane vortex. We can understand this if we consider that the in-plane vortex has a star shape, whose orientation changes with $\varphi$. As we will see below, a sixfold star vortex core whose shape is locked to the crystal lattice can provide such a behavior.

\begin{figure}[tb]
\centering
\includegraphics[width=0.9\columnwidth]{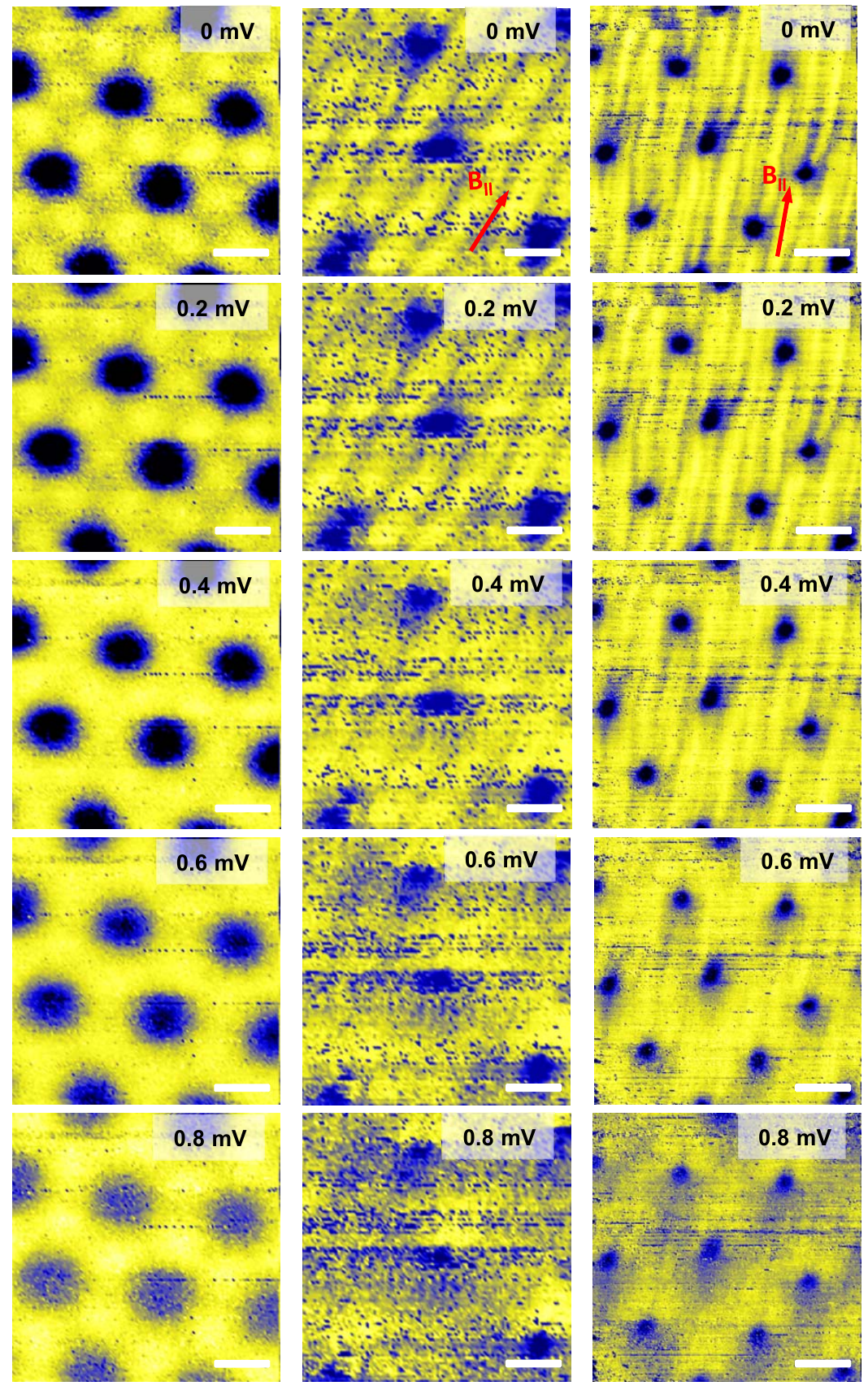}
\caption{\textbf{Tilted vortex cores vs bias voltage}. Conductance maps showing the evolution of the shape of the vortex cores as a function of the bias voltage. Left panels show the bias-voltage dependence of conductance images when the field of 0.6~T is parallel to the $c$ axis. Middle and right panels show the bias voltage dependence with nearly parallel fields ($\varphi=0$ middle panels and $\varphi=30^{\circ}$ right panels, $\theta \approx 90^{\circ}$ in both cases, see text and methods for uncertainty in $\theta$). We show schematically the direction of the tilt by the red arrows. Scale bars are of 30 nm in the left column and 120 nm in middle and right columns. Color scale is as in Fig.~\ref{f1}a.}
\label{f2}
\end{figure}

In Fig.~\ref{f2} we compare the bias voltage dependence in perpendicular and tilted fields. In perpendicular magnetic fields, when increasing the bias voltage, we observe the same behavior as reported previously, showing the star shaped DOS associated to the in-plane anisotropy of the superconducting gap \cite{Hess90,Guillamon08PRB}. In tilted magnetic fields, the bias voltage dependence of the conductance maps is completely modified. The stripes observed at zero bias are no longer seen at higher bias (middle and right panels in Fig.~\ref{f2}). Instead, we observe broad dark-bright patterns along the tilt of the magnetic field.

Note that the images show several outgoing vortices, in addition to the stripes. From the vortex density we obtain $\theta=87.5^{\circ}$ in Fig.~\ref{f1}b left panel (and in Fig.~\ref{f2} middle column), and $\theta=85.6^{\circ}$ in Fig.~\ref{f1}b right panel (and in Fig.~\ref{f2} right column). These angles are within the possible misorientation with respect to a parallel magnetic field that can be obtained in our experiment (see methods). As we discuss in detail below, vortices might come out either perpendicular or at an angle to the surface. The shape of the vortices we observe in highly tilted fields, as well as their bias voltage dependence, is totally different than the shape of vortices in perpendicular fields. Their structures actually merge into the stripes when leaving the vortex centers, having the same two vs three fold structure of rays that extend into the stripes far from vortex cores. They are also, as remarked above, considerably larger than vortices in perpendicular fields. Thus, we are observing vortices exiting the surface at an angle. 

\begin{figure}[tb]
\centering
\includegraphics[width=0.9\columnwidth]{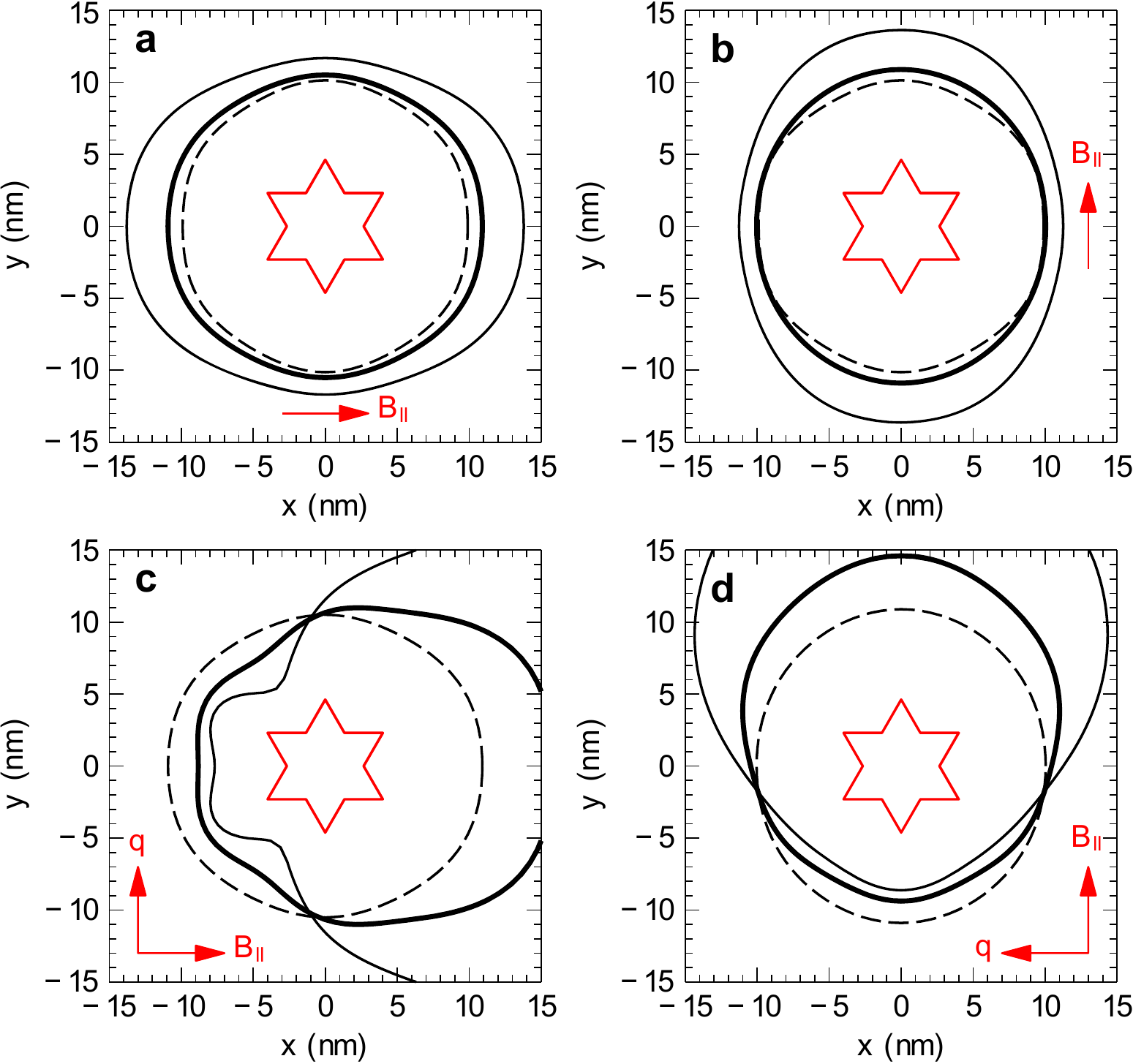}
\caption{\textbf{Order parameter distribution for tilted vortex cores}. Contours at a constant value of the order parameter are given for three values of the tilt of the magnetic field $\tau$ ($\tau=0$: dashed; $0.5$: thick; 1: thin) with (\textbf{a}) field tilted along the $a$ axis ($\varphi=0^{\circ}$) and (\textbf{b}) field tilted perpendicular to the $a$ axis ($\varphi=30^{\circ}$). (\textbf{c}) and (\textbf{d}) Results at $\tau=0.5$ for three values of $q$ ($qa=0$: dashed; $0.004$: thick; $0.008$: thin, $a$ being the lattice parameter) for $\varphi=0^{\circ}$ and $30^{\circ}$, respectively. The shape of the vortex core from the zero bias tunneling conductance in perpendicular fields is schematically marked by the red star for reference.}
\label{f3}
\end{figure}

To model such a situation, we perform microscopic calculations of tilted vortex cores in a superconductor with a sixfold gap anisotropy. It turns out that it is a formidable task to calculate the vortex core shape in tilted magnetic fields because one needs to obtain the DOS as a function of the position in three-dimensional space with a high spatial accuracy and then analyze the magnetic field and gap structure as a function of the tilt. This has never been achieved, to our knowledge. We start by building DOS maps of isolated vortices in an effective two-dimensional model and calculate the DOS maps as a function of energy and the azimuthal angle of the magnetic field. We simulate a vortex tilted to the hexagonal plane by a two-fold distribution of the supercurrent along the tilt, using calculations of Ref.~\cite{PhysRevB.46.366}. We introduce in addition a shift $q$ in the condensate momentum to account for the current distribution due to the in-plane vs out-of-plane crystalline anisotropy. In Fig.~\ref{f3} we show the vortex-core shape from contours of constant order parameter around the vortex cores when tilting the magnetic field (a and b) with zero $q$ and with nonzero $q$ (c and d). With zero $q$, we see that the DOS pattern becomes elliptical. With a finite $q$ we find that the DOS map of the vortex core develops a comet like shape, increasing considerably its size along the tilt of the magnetic field. The comet shape has a two-fold structure perpendicular to the tilt when the tilt is along a crystalline axis ($\varphi=0^{\circ}$) and a triangular form when the tilt is at $30^{\circ}$ to a crystalline axis ($\varphi=30^{\circ}$).

\begin{figure}[tb]
\centering
\includegraphics[width=0.9\columnwidth]{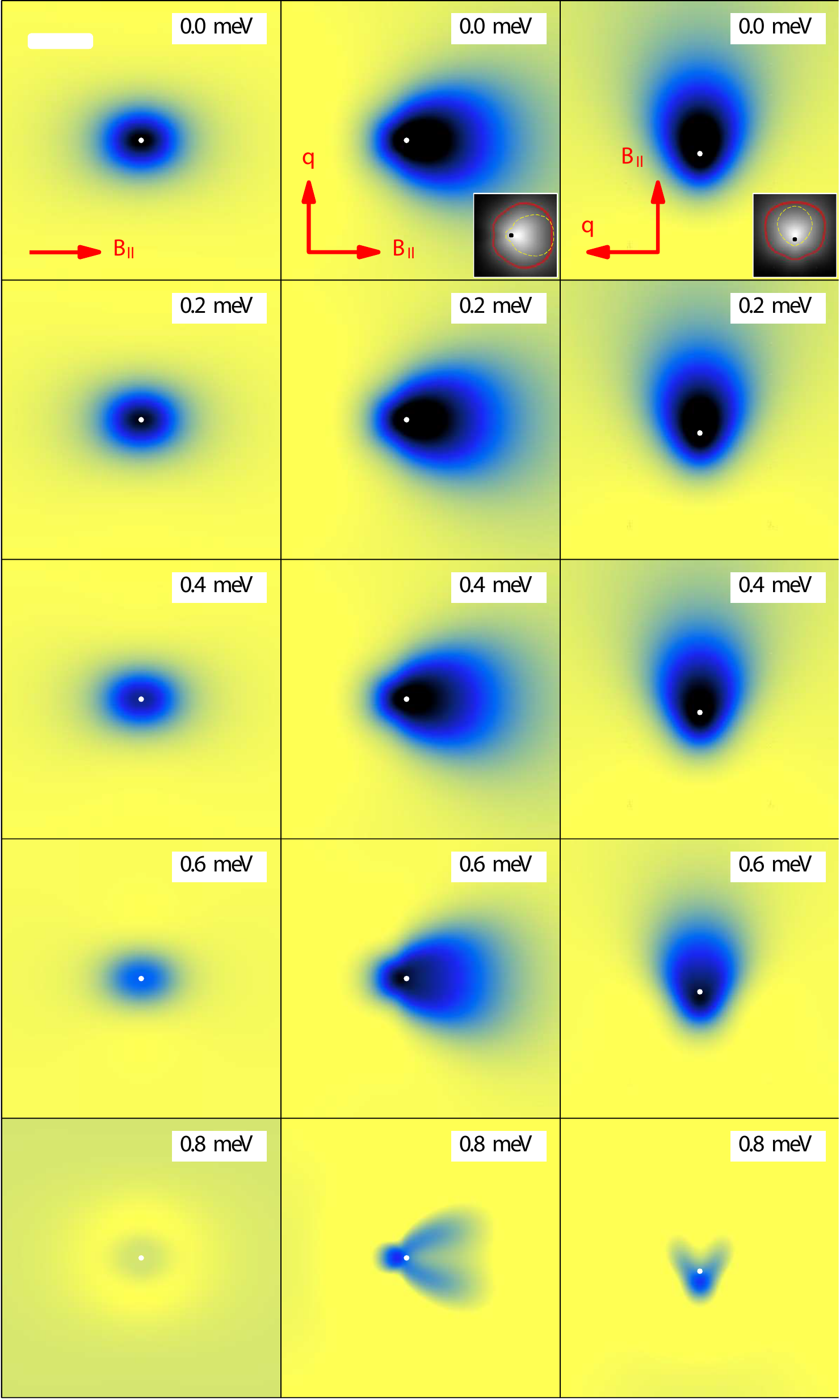}
\caption{\textbf{Calculated tilted vortex core shape vs energy}. In the left column we show the DOS at the indicated energies for a vortex with field tilted along the $a$ axis ($\varphi=0^{\circ}$) without taking into account screening surface currents, as in Fig.~\ref{f3}a. In the other columns we show the result for $\varphi=0^{\circ}$ (middle column) and $\varphi=30^{\circ}$ (right column) with a current of $qa=0.008$ in the direction shown by the red arrow in upper panels. The scale bar is of 30~nm. The color map goes from yellow (minimum DOS) to the zero bias peak in the DOS (as discussed in the supplementary information) in black. The white dots show the phase singularity point. The tilt strength is $\tau=0.5$ in all panels. Insets on top panels display the core shape on a logarithmic color scale together with the shape of the order parameter as in Fig.~\ref{f3}c and \ref{f3}d (dashed lines), and an iso-contour of the DOS (red lines).}
\label{f4}
\end{figure}

In Fig.~\ref{f4} we show the energy dependence of the DOS with the field tilted along the nearest-neighbor direction of the vortices and perpendicular to this direction (middle and right columns, respectively). We see that the vortex core shape obtained from the DOS is highly energy dependent. The spatial extension of the DOS around a vortex core is considerably larger in tilted fields. This is consistent with the observed increase in the vortex core size in the tunneling conductance when comparing maps made at perpendicular field with maps at nearly parallel fields (Fig.~\ref{f1}a). For energies close to the gap edge, at 0.8 meV the spatial extension is considerably reduced and has more structure.

Note that the model addresses a single vortex. Stripes result from vortex cores lying in-plane that reach the surface. Stripes of neighboring subsurface vortices extend to the surface and join each other. The shape of the order parameter extends along directions that change with the azimuthal angle $\varphi$ (Fig.~\ref{f3}), indicating that along these directions states of neighboring vortices overlap and lead to the observed stripes. In Fig.~\ref{f3}c we observe that the vortex extends preferentially along its sides, leading to the two-fold structure seen in the left panel of Fig.~\ref{f1}b. In Fig.~\ref{f3}d (Fig.~\ref{f1}b, right panel), we observe instead that the vortex core extends along the tilt and its sides, leading to the additional stripe exiting from the center of vortex cores in the right panel of Fig.~\ref{f1}b.

Note also that the model reproduces features of the bias voltage dependence of vortices exiting the surface at an angle. In particular, the increase of the DOS along the tilt for zero bias reverses for high bias. We can take in Fig.~\ref{f4} the phase singularity (white point) as a reference to compare DOS maps for different bias. The tunneling conductance is high (bluish, high DOS) at low bias, but lower (yellowish, low DOS) at high bias along the direction of the magnetic field. Similarly, the vortices observed in Fig.~\ref{f2} (middle and right columns) change their two-fold shape along the magnetic field.

\section*{Discussion and conclusions}

Vortices in tilted fields close to the surface are very different in 2H-NbSe$_2$ than in other materials. In the isotropic superconducting material {$\beta-$Bi$_2$Pd}, vortices at the surface have the same shape whatever the in-plane direction of the magnetic field \cite{Herrera17}. On the other hand, highly anisotropic materials like cuprates show two-dimensional pancake vortices, with the same shape at the surface than for perpendicular fields \cite{PhysRevB.70.184518,KirtleyKogan}. In both cases, the vortex core in tilted magnetic fields just shows the anisotropic properties in the plane, either because vortices bend close to the surface in {$\beta-$Bi$_2$Pd} or because vortices are fully confined within the layers (pancake vortices) in the cuprates. In 2H-NbSe$_2$ the observed DOS patterns are spatially highly structured at all bias voltages, indicating a more intricate situation.

In order to explain this situation, let us consider the energetics of vortices close to the surface in tilted magnetic fields and how this influences the three different systems mentioned here. We have to consider the balance between the elastic energy cost associated with vortex bending, which favors straight vortices exiting at an angle to the surface, and the cost associated with establishing strongly distorted current loops around vortices close to the surface which favors circular current loops and vortices exiting perpendicular to the surface \cite{Brandt95,KirtleyKogan}.

The shear modulus $c_{44}$  is given by $c_{44}\approx (1+k^2\lambda^2)^{-1}$ with $k$ being the wavevector for the distortion \cite{Brandt95}. Close to the surface, the wavevector relevant for this problem is $k=1/a$ where $a$ is the bending radius. $a$ is given by the minimum between the intervortex distance $a_0$ and the penetration depth $\lambda$ \cite{Brandt93}. For the magnetic fields considered here, $a_0$ is of about 90 nm, always smaller than $\lambda$, thus $a\approx a_0$. On the other hand, the current distribution surrounding the vortex is determined by the anisotropy of the upper critical field and is elliptical, with its short axis along the direction where the upper critical field is higher (and the coherence lengths smaller) \cite{Brandt95,Blatter94,PhysRevB.42.2631}.

In $\beta-$Bi$_2$Pd, vortices exiting the surface have a circular shape and bend close to the surface \cite{Herrera17}. The shear modulus of the vortex lattice is isotropic and the lattice is sufficiently soft that it is energetically favorable to exit perpendicular to the surface for all tilts. The upper critical field in $\beta-$Bi$_2$Pd is low (0.6~T) and the coherence length of about 20 nm, so that vortices are larger than in 2H-NbSe$_2$ and the cuprates and certainly round because the upper critical field is nearly isotropic \cite{PhysRevB.92.054507,PhysRevB.93.144502}. In $\beta-$Bi$_2$Pd, $\lambda\approx$ 100 nm, which increases the elastic energy associated with vortex bending close to the surface as compared to materials with larger $\lambda$ \cite{Herrera17,PhysRevB.93.144502}. However, this is balanced by the gain in establishing current loops parallel to the surface with bent vortices \cite{PhysRevB.46.366}. 

In the cuprates, tilted fields create arrangements of crossing pancake and Josephson vortex lattices whose structure strongly depends on the temperature, magnetic field and tilt \cite{Vlasko15,PhysRevLett.83.187,PhysRevB.91.014516,Bending1999}. Lattices of pancake vortices slightly shifted between layers can appear under some conditions, leading in essence to tilted lines of vortices that are expected to exit at an angle to the surface \cite{PhysRevB.46.366,PhysRevLett.88.237001,PhysRevB.48.1180,Tonomura2001}. However, at least for materials in which the $c$-axis coherence length is below the interlayer distance, there is no reason for vortex bending as no Abrikosov vortices are established between layers. Current loops of Abrikosov vortices are parallel to the surface, and the balance is rather played by the interaction between Abrikosov and Josephson vortex lattices.

Here in 2H-NbSe$_2$, the upper critical field anisotropy is of a factor of three and the out of plane coherence length is well above the $c$-axis lattice constant \cite{Foner73,Nader14,Hess94}. This creates elliptical Abrikosov vortices (there are no Josephson vortices) with the short axis perpendicular to the surface. This helps establishing current loops close to the surface, as the vortex is anyhow shorter along the direction perpendicular to the surface. Vortices can thus exit at an angle to the surface, neither parallel nor perpendicular to the surface. This leads to an intricate surface current pattern that modifies the vortex core shape at all bias, as shown by the vortices imaged in our experiment.

On the other hand, the appearance of stripes is also unique to 2H-NbSe$_2$. As we show above, this is due to subsurface vortex cores. The zero bias peak due Caroli--de Gennes--Matricon states is indeed a characteristic feature of 2H-NbSe$_2$ and it extends indeed over distances far from the vortex core \cite{Hess89,Guillamon08PRB,Hess90}. In our experiments, the distance between stripes from subsurface vortices is comparable to the intervortex distance expected for the applied in-plane magnetic field, with each subsurface vortex providing a different number of stripes, depending on the in-plane direction of the tilt, as shown in Fig.~\ref{f1}b and as discussed in connection with Fig.~\ref{f3}.

The combined appearance of stripes and vortices could be also due, in principle, to crossing perpendicular vortex and parallel vortex lattices\cite{PhysRevLett.70.2948}. Within such a situation, however, it seems quite difficult to obtain patterns of stripes with distances that change when varying the in-plane direction of the tilt.

We can conclude that the differences between 2H-NbSe$_2$, $\beta-$Bi$_2$Pd and the cuprates are due to a totally different DOS pattern for parallel magnetic fields. The pattern results from subsurface vortex cores and provides access to the shape of in-plane vortices and thus to the gap anisotropy in the out-of-plane three-dimensional part of the Fermi surface.

The shape of the superconducting gap along the $c$ axis should influence the dependence of the core shape with the in-plane direction of the tilt. As we show above, the main asymmetric features observed in our experiment and their dependence with the azimuthal angle can be explained by a sixfold anisotropic superconducting gap in 2H-NbSe$_2$. This shows that the sixfold gap anisotropy is spread over the whole Fermi surface. 

It is useful to compare to the situation in MgB$_2$, where vortex cores have been studied for magnetic fields always applied perpendicular to the surface, but with surfaces out- as well as in-plane \cite{PhysRevLett.89.187003,PhysRevB.68.100508}. An elongated vortex core shape was expected for fields in-plane, due to the two-dimensional part of the Fermi surface \cite{PhysRevB.70.144508}. This was however not observed. Vortices are round out- and in-plane, with their shape being dominated by the three-dimensional part of the Fermi surface, both the vortex core shape in the DOS \cite{PhysRevLett.89.187003,PhysRevB.68.100508,PhysRevB.70.144508} as well as in the magnetic field pattern \cite{PhysRevB.91.054505}. It was concluded that tunneling is dominated in all cases by the three-dimensional part of the Fermi surface. Tunneling in 2H-NbSe$_2$ is also dominated by the three dimensional part of the Fermi surface, here due to the stronger contribution to tunneling of the Se orbitals \cite{Johannes06}. However, the situation is quite different. Whereas the three-dimensional pocket amounts for the largest part of the DOS at the Fermi level in MgB$_2$, the Se pocket in 2H-NbSe$_2$ is small and amounts just for 20\% of the DOS \cite{Johannes06}. The gap anisotropy in  2H-NbSe$_2$ was until now just associated to the Nb two dimensional sheets. As we show here, the sixfold gap anisotropy remains out-of-plane, which suggests that it is present over the whole Fermi surface.

Interband interactions are strong in 2H-NbSe$_2$ \cite{Johannes06}. Although there are evidences for a two gap behavior from tunneling spectroscopy, penetration depth and thermal conductivity \cite{Boaknin03,Guillamon08PRB,Fletcher07,Johannes06,Rodrigo04c}, the obtained spread in gap values is of a factor of 1.5 or at most 2. Our measurement suggest that interactions are such that the sixfold anisotropy in the shape of the superconducting gap is inherited from the Nb two-dimensional bands over to the whole Fermi surface.

STM in tilted magnetic fields is thus a sensitive probe of the superconducting electronic DOS, particularly in three-dimensional superconductors with strongly anisotropic and sharply defined vortex core bound states. The main advantage with respect to studying different crystalline surfaces is that the direction of the magnetic field can be varied at will on a single surface. Efforts, still under development, have been also devoted to other techniques. Quasiparticle interference has been extended out of the surface plane by triangulating electronic wavefunctions from in-depth impurities or studying inter and intraband scattering \cite{Hoffman11,Petersen99,Hoffman02a,Simon11,Weismann09,Franke15}. Macroscopic specific heat and thermal conductivity measurements in rotating fields have been also performed with the aim to obtain the anisotropy of the superconducting gap \cite{0953-8984-18-44-R01}. These provide the spatially averaged DOS and are influenced by scattering effects, whereas STM provides directly the spatially resolved electronic DOS, i.e. in and around vortex cores.

\section*{Methods}
\textbf{Model and calculations.} We create a microscopic model with a sixfold anisotropic superconducting gap, by a triangular lattice and a tight-binding dispersion (see the supplementary information for more information). To calculate the spatial dependence of the superconducting DOS we need a very large system size. In the low-energy region the inter-level spacing is $\Delta E\sim\hbar v_{\mathrm{F}}\Delta k$ with $\Delta k=2\pi/(Na)$, where $Na$ is the linear system size. In order to reach a resolution $\Delta E\lesssim1$~meV of the order of the superconducting gap, a total number of unit cells $N^2\gtrsim 200\,000$ is required. This sets the size of the Bogoliubov-de Gennes Hamiltonian to at least $400\,000\times400\,000$. Straight diagonalization is therefore not an option. Instead we use the method described in Ref.~\cite{Covaci10}, the DOS is expanded on Chebyshev polynomials of the Hamiltonian and can thus be evaluated iteratively with low memory cost, even for very large systems. We use a finite lattice made of $M$ concentric hexagons surrounding the site where the DOS is calculated. To reach the desired accuracy we perform the calculation with $M=1000$, corresponding to a lattice of 3\,003\,001 sites. The Chebyshev expansion is truncated at order $4M$ and terminated using the Jackson kernel \cite{Weisse06,Berthod16}. To obtain results in an applied magnetic field and as a function of position and energy we use a Lawrence-Doniach model following Ref.~\cite{PhysRevB.46.366}. We are not aware of an equivalent model for vortex lattices---hence we consider only isolated vortices. In the supplementary information we solve the nonlinear model of Ref.~\cite{PhysRevB.46.366} to leading order in the tilt angle and get an analytical expression for the phase. The distortion is proportional to the tilt strength defined as $\tau=\xi_{ab}^2/(2\xi_{c}^2)\tan^2\theta$, with $\theta$ being the polar angle. The DOS maps and anisotropies of the order-parameter are determined self-consistently as a function of $\tau$.  We use tilt strengths of $\tau\approx 0.5$, which correspond to small $\theta\approx 15^{\circ}$. We believe that this accounts well for our situation, with a three dimensional, strongly anisotropic superconductor. To further address this anisotropy, we introduce an additional phase in the order parameter, i.e., $\Phi(\vec{r})\to\Phi(\vec{r})-\vec{q}\cdot\vec{r}$, with $q$ being a surface current. We take $q$ up to about one third of the value corresponding to the critical current.\\

\textbf{Scanning tunneling microscopy methods.} We use a setup described in Ref.~\cite{Galvis15} which consists of a dilution refrigerator with a STM thermally anchored to the mixing chamber reaching temperatures of 150~mK. The STM is located at the center of a three axis magnet. The three axis magnet consists of a long solenoid providing the $z$-axis magnetic field and a pair of split coils for $x$- and $y$-axis magnetic field components. We estimate the accuracy in the determination of the azimuthal and polar angles $\theta$ and $\varphi$ at the magnetic fields used here (of order of a Tesla) to be about 4--5$^{\circ}$. This uncertainty is composed of a possible slight misalignment of the surface of the sample with respect to the coil system and of remanent magnetic fields inside the coils, that can be in the range of the mT. We measure high quality 2H-NbSe$_2$ samples grown using the usual iodine vapor transport method. The in-plane tilt direction is corrected to obtain values for $\varphi$ that are fixed to the atomically resolved hexagonal Se lattice obtained by STM imaging (see also Fig.~\ref{fs3}). We take tunneling conductance maps as a function of the bias voltage, as usual in vortex imaging using STM. Sample surface has been prepared by cleaving and the Au tip is prepared using a pad of Au as described in Ref.~\cite{Suderow11}.


\section*{Acknowledgments}

We acknowledge discussions with V.G. Kogan and J.R. Kirtley. Work done in Madrid was supported by the Spanish Ministry of Economy and Competitiveness (FIS2014-54498-R, MDM-2014-0377), by the Comunidad de Madrid through program Nanofrontmag-CM (S2013/MIT-2850), by EU (European Research Council PNICTEYES grant agreement 679080, FP7-PEOPLE-2013-CIG 618321 and COST Action CA16218) and by Axa Research Fund. SEGAINVEX-UAM is also acknowledged. We also acknowledge the support of Departamento Administrativo de Ciencia, Tecnología e Innovación, COLCIENCIAS (Colombia) Programa Doctorados en el Exterior Convocatoria 568-2012 and the Cluster de investigaci\'on en ciencias y tecnolog\'ias convergentes de la Universidad Central (Colombia). Work done in Geneva was supported by the Swiss National Science Foundation under Division II. Calculations were done in the University of Geneva with the clusters Mafalda and Baobab.

\onecolumngrid
\newpage
\begin{center}

{\large\textbf{\boldmath
Supplementary Information\\ [0.5em] {\small for} \\ [0.5em]
Tilted vortex cores and superconducting gap anisotropy in 2H-NbSe$_2$
}}\\[1.5em]

J.A. Galvis,$^{1, 2}$ E. Herrera,$^{1, 3}$ C. Berthod,$^4$ S. Vieira,$^1$ I. Guillam{\'o}n,$^1$ and H. Suderow$^1$\\[0.5em]

\textit{\small
$^1$\!{Laboratorio de Bajas Temperaturas y Altos Campos Magn\'eticos,\\ Unidad Asociada UAM/CISC, Departamento de F\'isica de la Materia Condensada,\\ Instituto de Ciencia de Materiales Nicol\'as Cabrera,\\ Center for Condensed Matter Physics (IFIMAC),\\ Universidad Aut\'onoma de Madrid, E-28049 Madrid, Spain}\\
$^2$\!{Departamento de ciencias naturales, Facultad de ingenieria y ciencias b{\'a}sicas, Universidad Central, Bogot\'a 110311, Colombia}\\
$^3$\!{Departamento de F\'isica, Universidad Nacional de Colombia, Bogot\'a 111321, Colombia}\\
$^4$\!{Department of Quantum Matter Physics, University of Geneva,\\ 24 quai Ernest-Ansermet, 1211 Geneva, Switzerland}
}

\vspace{2em}
\end{center}

\twocolumngrid
\renewcommand{\thefigure}{S\arabic{figure}}
\setcounter{figure}{0}

We provide the mathematical derivations used to establish our model and present results of the model for zero magnetic field and for perpendicular fields. We further discuss additional features of the experiment, such as atomic resolution images of the surface and tunneling conductance curves within and around stripes. We also discuss the changes in the vortex lattice structure when tilting the magnetic field. Finally, we discuss a pattern of stripes for azimuthal angles in between those discussed in the main text.

\section*{Model}

We have chosen the simplest possible account of tilted vortices in 2H-NbSe$_2$. We need to reproduce the results in perpendicular fields and then calculate the DOS for tilted fields. The simplest possible way to reproduce the results in perpendicular fields is to model the tubular Fermi surface sheets within a single layer. This provides a six-fold anisotropic superconducting gap. It turns out that solving the Bogoliubov--de Gennes equations for three-dimensional vortex cores in tilted fields and in an anisotropic superconductor is prohibitively difficult. We thus chose to take the mentioned single layer with a sixfold gap anisotropy and calculate the DOS obtained by tilting the magnetic field out of the layer. As we show below, we find solvable equations for not too high tilts out of the layer. As we discuss in the main text, our results suggest that this procedure accounts for some relevant features of vortex cores in an anisotropic three-dimensional superconductor in parallel magnetic fields.

\subsection*{Superconducting density of states and maps for perpendicular magnetic fields}

\begin{figure}[tb]
\centering
\includegraphics[width=0.95\columnwidth]{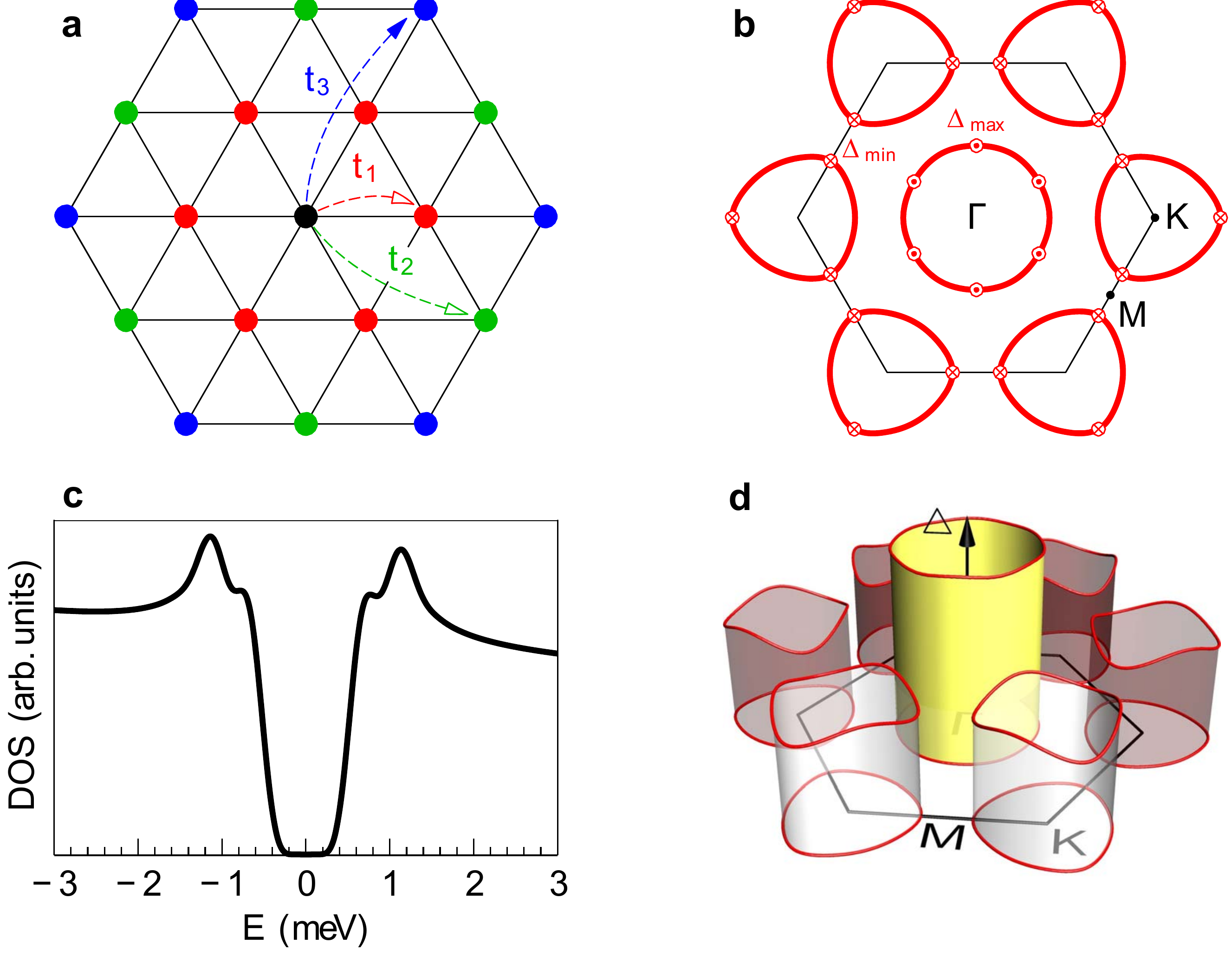}
\caption{
(\textbf{a}) Hopping amplitudes to the first (red), second (green), and third (blue) neighbors on the triangular lattice. (\textbf{b}) Hexagonal Brillouin zone and Fermi surface of the model; dots and crosses indicate the locations of maxima and minima of the superconducting gap, respectively. (\textbf{c}) Zero-field superconducting DOS. (\textbf{d}) Perspective view of the superconducting gap along the Fermi surface.}
\label{fs1}
\end{figure}

In Fig.~\ref{fs1}a we provide a schematic view of the microscopic input used for our tight-binding model. In phonon-mediated superconductors like 2H-NbSe$_2$, the superconducting properties are set by interactions taking place in the immediate vicinity of the Fermi surface. The relevant energy range is determined by the Debye scale which is $\hbar\omega_{\mathrm{D}}=18$~meV for 2H-NbSe$_2$ \cite{s_Bevolo74}. We include hopping amplitudes $t_{1,2,3}$ up to the third neighbors. The corresponding electronic band dispersion measured from the chemical potential $\mu$ is
	\begin{multline}\label{eq:band}
		\xi_{\vec{k}}=2t_1\left[\cos(k_xa)+2\cos(k_xa/2)\cos(\sqrt{3}k_ya/2)\right]\\
		+2t_2\left[2\cos(3k_xa/2)\cos(\sqrt{3}k_ya/2)+\cos(\sqrt{3}k_ya)\right]\\
		+2t_3\left[\cos(2k_xa)+2\cos(k_xa)\cos(\sqrt{3}k_ya)\right]-\mu.
	\end{multline}
$a=3.45$~\AA\ is the in-plane lattice parameter. For simplicity we use a one-band model. The Fermi-surface topology ($\xi_{\vec{k}}=0$) only fixes the ratios $t_2/t_1$, $t_3/t_1$, and $\mu/t_1$. We take the first two ratios from Ref.~\cite{s_Inosov08} by averaging the values reported for the two bands and fix the third ratio to match the Fermi surface at best. The remaining parameter $t_1$ sets the Fermi velocity, which in turn controls the superconducting coherence length. We use $(t_1,t_2,t_3,\mu)=(5.6,11.9,2.5,-9.4)$~meV. The average Fermi velocities are $v_{\mathrm{F}}^{\Gamma}=0.36\times10^7$~cm/s on the Fermi-surface pocket centered on the $\Gamma$ point and $v_{\mathrm{F}}^{\mathrm{K}}=0.25\times10^7$~cm/s on the pockets surrounding the K points. The phonon-mediated pairing interaction is represented by a BCS model involving both onsite interaction $V_0$ and nearest-neighbor interaction $V_1$ with an energy cutoff at $\pm\hbar\omega_{\mathrm{D}}$. To obtain a superconducting order parameter that varies in-plane we define two parameters, $\Delta_0=-V_0\langle\psi_{\vec{r}\uparrow}\psi_{\vec{r}\downarrow}\rangle$ describing interactions for two electrons on the same site and $\Delta_1/6=-V_1\langle\psi_{\vec{r}\uparrow}\psi_{\vec{r}'\downarrow}\rangle$ for two electrons on neighboring sites. We find
	\begin{equation}\label{eq:Deltak}
		\Delta_{\vec{k}}=\Delta_0+\frac{\Delta_1}{3}\left[\cos(k_xa)+2\cos(k_xa/2)\cos(\sqrt{3}k_ya/2)\right].
	\end{equation}
For onsite and nearest-neighbor interactions having opposite signs ($V_0<0$ and $V_1>0$), $\Delta_0$ and $\Delta_1$ have the same sign, as expected in 2H-NbSe$_2$. To set their values we require that the smallest and largest gaps on the Fermi surface are 0.56 and 1.04~meV, respectively, like in the empirical expression $\Delta_0(1+c_A\cos6\vartheta)$ where $\Delta_0=0.8$~meV and $c_A=0.3$ \cite{s_Hayashi97}. The resulting zero-field DOS is displayed in Fig.~\ref{fs1}c.

\subsection*{Isolated vortex and vortex lattice in perpendicular field}

\begin{figure}[tb]
\centering
\includegraphics[width=0.9\columnwidth]{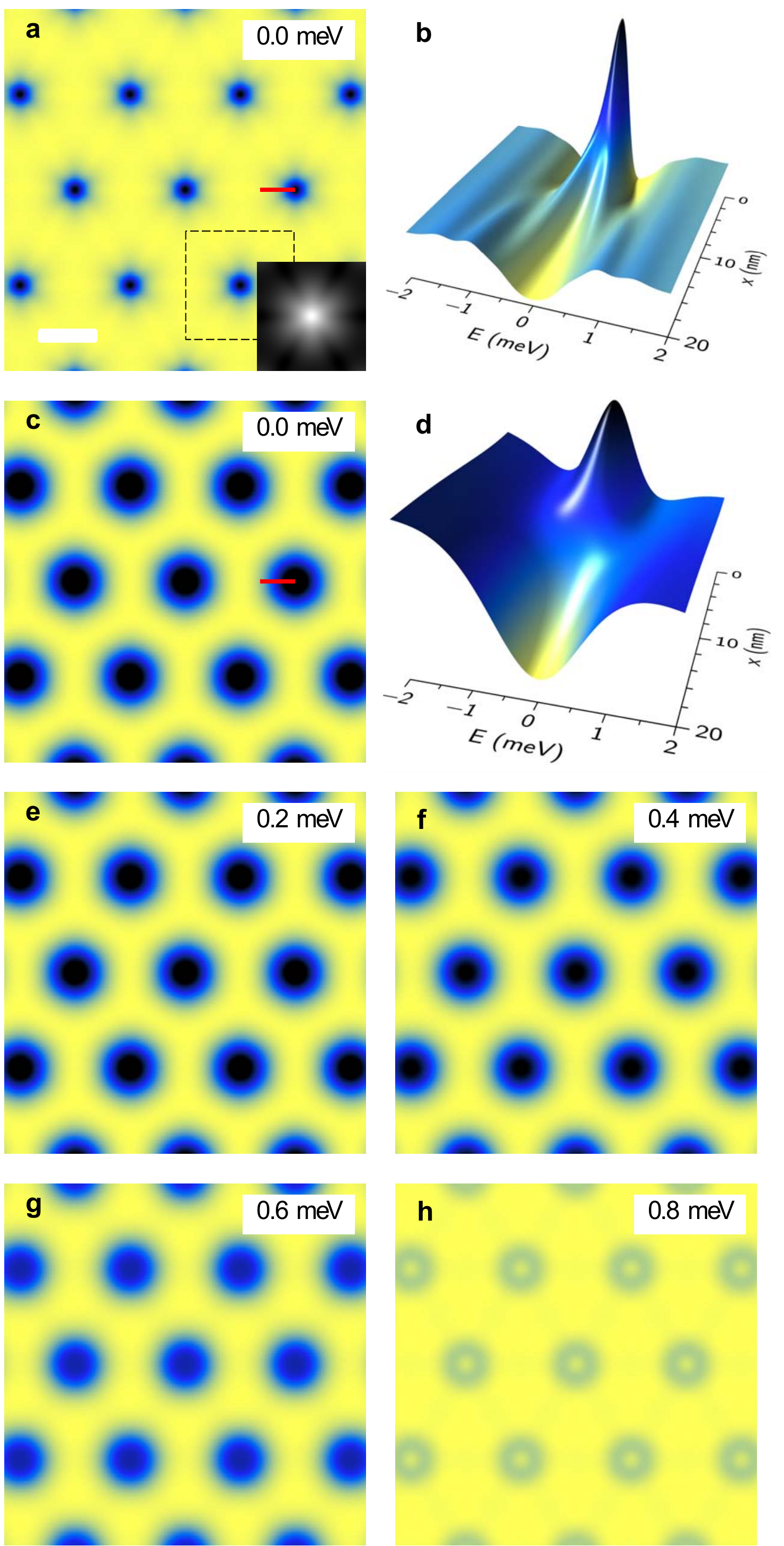}
\caption{(\textbf{a}) Zero-bias DOS and (\textbf{b}) energy-dependent DOS along the red line drawn in (\textbf{a}). The inset in (\textbf{a}) shows one core on a logarithmic color scale, highlighting the star shape pointing toward next-nearest neighbor vortices. In (\textbf{c})--(\textbf{h}) the DOS is broadened with a Lorentzian of width 0.7~meV. (\textbf{c}) and (\textbf{d}) are like (\textbf{a}) and (\textbf{b}) with the broadened DOS. (\textbf{e})--(\textbf{h}) show the broadened DOS at finite energies. The scale bar is of 30~nm. The color scale in (\textbf{c}) and (\textbf{e})--(\textbf{h}) goes from yellow in-between vortices to a common maximum value in black.}
\label{fs2}
\end{figure}

We introduce the magnetic field via the Peierls substitution. In Fig.~\ref{fs2}a we show the resulting DOS map at zero bias and in Fig.~\ref{fs2}b the DOS along a path running from the vortex core along the $a$ axis. We reproduce the well known structure of the DOS due to Caroli--de Gennes--Matricon states \cite{s_Hayashi97,s_Hess89,s_Guillamon08PRB,s_Hess90}. To highlight the main features and better compare with experiment, we introduce a Lorentzian smearing of the DOS (Fig.~\ref{fs2}d). The evolution of the resulting DOS maps with energy is displayed in Figs.~\ref{fs2}e--\ref{fs2}h.

For a magnetic field oriented along the $c$ axis we start with the Ansatz
	\begin{multline}\label{eq:vortex}
		\Delta_{\vec{r}\vec{r}'}=\delta_{\vec{r}\vec{r}'}\Delta_0
		\tanh\left(\frac{r}{\xi(\vartheta_{\vec{r}})}\right)e^{-i\vartheta_{\vec{r}}}\\
		+\delta_{|\vec{r}-\vec{r}'|,a}\frac{\Delta_1}{6}\tanh\left(\frac{|\vec{r}+\vec{r}'|/2}
		{\xi\big(\vartheta_{\frac{\vec{r}+\vec{r}'}{2}}\big)}\right)
		e^{-i\vartheta_{\frac{\vec{r}+\vec{r}'}{2}}}.
	\end{multline}
This provides a sixfold anisotropy in the DOS maps that can be parameterized by a length scale varying with the in-plane angle $\vartheta$ as
\begin{equation}\label{eq:star}
		\xi(\vartheta)={\textstyle\frac{1}{2}}(\xi_{10}+\xi_{01})+{\textstyle\frac{1}{2}}(\xi_{10}-\xi_{01})
		\cos(6\vartheta).
	\end{equation}
Here $\xi_{10}$ and $\xi_{01}$ are length scales describing the spatial extension of the DOS along the $(10)$ and $(01)$ directions. This Ansatz provides the DOS maps with just two parameters $\xi_{10}$ and $\xi_{01}$ that are determined self-consistently under the constraint that $(\xi_{10}+\xi_{01})/2=10$~nm. The constraint was implemented by adjusting the hopping amplitude $t_1$. In a magnetic field of 0.6~T we find only a slight anisotropy, which is enough to discuss and compare with our experiments.

Recently, a functional form more precise than tanh was established for the spatial dependence of the DOS in vortices that are in-plane isotropic \cite{s_Fente16}. Here, vortex cores are anisotropic and the anisotropy is not influenced by using more involved functional forms than tanh. We thus stick to the most simple function giving a spatially dependent DOS.

\subsection*{Isolated vortex in tilted field}

We are now equipped to calculate maps of the DOS for perpendicular magnetic fields. To make the calculations in tilted fields, we start by solving the model of Ref.~\cite{s_Bulaevskii92} for pancake vortices. The equations for the order-parameter phase $\Phi_n(\vec{r})$ in the layer $n$ are expressed in terms of a function $\varphi_{\vec{k}}$ which is the two-dimensional Fourier transform of the phase difference between successive layers. For the case of a single vortex tilted by an angle $\theta$ from the vertical in the $(x,z)$ plane this function satisfies the nonlinear equation
	\begin{equation}
		k^2\varphi_{\vec{k}}+\left(\frac{Q^2}{\Gamma^2}+\frac{1}{\lambda_c^2}\right)W_{\vec{k}}
		=4\pi i\frac{k_y}{k_x}\sin\left(\frac{k_xa}{2}\right).
	\end{equation}
$a=s\tan\theta$ with $s$ the interlayer spacing, $\Gamma=\xi_{ab}/\xi_{c}$, $Q^2=(2/s^2)[1-\cos(k_xa)]$, and $W_{\vec{k}}=[\sin\varphi(\vec{r})]_{\vec{k}}$ is the Fourier transform of $\sin\varphi(\vec{r})$. The derivatives of $\Phi_n(\vec{r})$ are denoted $\Phi_{n,\alpha}=\partial_{\alpha}\Phi_n(\vec{r})$ and their three-dimensional Fourier transform
	\begin{equation}
		\Phi_{\alpha}(\vec{k},q)=\sum_n\int d^2rdz\,\Phi_{n,\alpha}(\vec{r})\delta(z-ns)
		e^{-i(\vec{k}\cdot\vec{r}+qz)}
	\end{equation}
are related to the function $\varphi_{\vec{k}}$ by
	\begin{subequations}\begin{align}
		2\Phi_x(\vec{k},q)\sin\left(\frac{k_xa}{2}\right)&=k_x\varphi_{\vec{k}}\delta(qs+k_xa)\\
		\nonumber
		2\Phi_y(\vec{k},q)\sin\left(\frac{k_xa}{2}\right)&=\left[k_y\varphi_{\vec{k}}
		-\frac{4\pi i}{k_x}\sin\left(\frac{k_xa}{2}\right)\right]\\ &\quad\times\delta(qs+k_xa).
	\end{align}\end{subequations}
We solve this nonlinear system of equations in the limit of a small tilt angle, such that the phase difference between successive layers is small and $\sin\varphi(\vec{r})\approx\varphi(\vec{r})$. We then obtain
	\begin{equation}
		\Phi_x(\vec{k},q)=\frac{2\pi i k_y}{k^2+2\lambda_{\mathrm{J}}^{-2}[1-\cos(k_xa)]+\lambda_c^{-2}}
		\delta(qs+k_xa)
	\end{equation}
with $\lambda_{\mathrm{J}}=s\xi_{ab}/\xi_{c}$. Fourier transforming back we have
	\begin{equation}
		\Phi_x(\vec{r},z)=i\int\frac{d^2k}{(2\pi)^2}\,\frac{k_ye^{i(\vec{k}\cdot\vec{r}-k_xz\tan\theta)}}
		{k^2+2\lambda_{\mathrm{J}}^{-2}[1-\cos(k_xa)]+\lambda_c^{-2}}.
	\end{equation}
The $k_y$ integral is given by the expression
	\begin{equation}
		\int_{-\infty}^{\infty}\frac{dk_y}{2\pi}\frac{k_ye^{ik_yy}}{A^2+k_y^2}
		=\frac{i}{2}\mathrm{sign}(y)e^{-A|y|},
	\end{equation}
which can be used in the limit of small angles where $k_x^2+2\lambda_{\mathrm{J}}^{-2}[1-\cos(k_xa)]+\lambda_c^{-2}>0$. We then take the limit $\lambda_c\to\infty$ and use the small-angle expansion
	\begin{multline*}
		e^{-\sqrt{k_x^2+2\lambda_{\mathrm{J}}^{-2}[1-\cos(k_xa)]}\,|y|}\\
		=e^{-|k_xy|}\left[1-\frac{a^2}{2\lambda_{\mathrm{J}}^2}|k_xy|+O(a^4)\right]
	\end{multline*}
to obtain
	\begin{multline}
		\Phi_x(\vec{r},z)=\frac{1}{2\pi}\left\{\frac{-y}{(x-z\tan\theta)^2+y^2}\right.\\
		\left.+\frac{a^2}{2\lambda_{\mathrm{J}}^2}
		\frac{-y[(x-z\tan\theta)^2-y^2)}{[(x-z\tan\theta)^2+y^2]^2}\right\}.
	\end{multline}
Since $\Phi_x(\vec{r},z)=\sum_n\Phi_{n,x}(\vec{r})\delta(z-ns)$ we can deduce the phase gradient in the layer $n=0$:
	\begin{equation}
		\partial_x\Phi_0(\vec{r})=\frac{-y}{x^2+y^2}\left(1+\frac{\xi_{c}^2}{2\xi_{ab}^2}
		\tan^2\theta\frac{x^2-y^2}{x^2+y^2}\right).
	\end{equation}
The gradient along $y$ follows in the same way:
	\begin{equation}
		\partial_y\Phi_0(\vec{r})=\frac{x}{x^2+y^2}\left(1+\frac{\xi_{ab}^2}{2\xi_{c}^2}
		\tan^2\theta\frac{x^2-y^2}{x^2+y^2}\right).
	\end{equation}
The solution for the phase is therefore
	\begin{equation}
		\Phi_0(\vec{r})=\arg(x+iy)+\frac{\xi_{ab}^2}{2\xi_{c}^2}\tan^2\theta\frac{xy}{x^2+y^2}.
	\end{equation}
Upon a change $y\to-y$ due to different gauge conventions in Ref.~\cite{s_Bulaevskii92} and in the present work, and allowing the magnetic field to be tilted in a plane forming an angle $\alpha$ with the $x$ axis, this gives the order-parameter phase
	\begin{multline}\label{eq:tilted}
		\Phi(\vec{r})=\arg(x-iy)\\
		-\tau\frac{(x\cos\alpha-y\sin\alpha)(x\sin\alpha+y\cos\alpha)}{x^2+y^2}.
	\end{multline}
The first term is equivalent to the phase $-\vartheta_{\vec{r}}$ of Eq.~(\ref{eq:vortex}). $\alpha$ is the direction of the tilt measured from the (10) direction, and $\tau=\xi_{ab}^2/(2\xi_{c}^2)\tan^2\theta$ measures the strength of the tilt, with $\xi_{ab}$ ($\xi_{c}$) the $c$-axis ($ab$-plane) coherence length and $\theta$ the tilt angle.

We substitute $\Phi(\vec{r})$ for $-\vartheta_{\vec{r}}$ in Eq.~(\ref{eq:vortex}) and generalize Eq.~(\ref{eq:star}) to allow for a twofold deformation along the direction $\alpha$:
	\begin{multline}\label{eq:6-oval}
		\xi(\vartheta)={\textstyle\frac{1}{2}}(\xi_{\parallel}+\xi_{\perp})
		+{\textstyle\frac{2}{3}}(\xi_{a}-\xi_{\perp})\cos[2(\vartheta-\alpha)]\\
		+{\textstyle\frac{1}{6}}(3\xi_{\parallel}-4\xi_{a}+\xi_{\perp})\cos[6(\vartheta-\alpha)].
	\end{multline}
Here $\xi_{\parallel}=\xi(\alpha)$ is the length scale for isocontours in the DOS in the direction of the tilt, $\xi_{\perp}=\xi(\alpha+\pi/2)$ in the direction normal to the tilt, and $\xi_{a}=\xi(\alpha+\pi/6)$ measures the sixfold anisotropy. Since the latter anisotropy is most likely locked with the microscopic lattice it may not develop at an angle of $\pi/6$ for all $\alpha$. We consider the cases $\alpha=0$ and $\alpha=\pi/2$ where it does. We determine the contours of constant order parameter, as displayed in Fig.~\ref{f3}a and \ref{f3}b.

In presence of a screening current, we also generalize Eq.~(\ref{eq:6-oval}) further to allow for an asymmetry between the front and back sides:
	\begin{multline}\label{eq:6-egg}
		\xi(\vartheta)={\textstyle\frac{1}{4}}(\xi^>_{\parallel}+2\xi_{\perp}+\xi^<_{\parallel})
		+{\textstyle\frac{1}{2}}(\xi^>_{\parallel}-\xi^<_{\parallel})\cos(\vartheta-\alpha)\\
		+{\textstyle\frac{2}{3}}\Big[\xi_{a}-\xi_{\perp}
		-{\textstyle\frac{\sqrt{3}}{4}}(\xi^>_{\parallel}-\xi^<_{\parallel})\Big]\cos[2(\vartheta-\alpha)]\\
		+{\textstyle\frac{1}{6}}\big[{\textstyle\frac{3}{2}}(\xi^>_{\parallel}+\xi^<_{\parallel})
		-4\xi_{a}+\xi_{\perp}+\sqrt{3}(\xi^>_{\parallel}-\xi^<_{\parallel})\big]\cos[6(\vartheta-\alpha)].
	\end{multline}
There are now four length scales, $\xi^>_{\parallel}=\xi(\alpha)$, $\xi^<_{\parallel}=\xi(\alpha+\pi)$, $\xi_{a}=\xi(\alpha+\pi/6)$, and $\xi_{\perp}=\xi(\alpha+\pi/2)$. We select a tilt strength $\tau=0.5$ and determine these four length scales self-consistently as a function of $q$. The results are displayed in Fig.~\ref{f3}c and \ref{f3}d of the main text.

\section*{The vortex and atomic lattices}

\begin{figure}[b]
\centering
\includegraphics[width=\columnwidth]{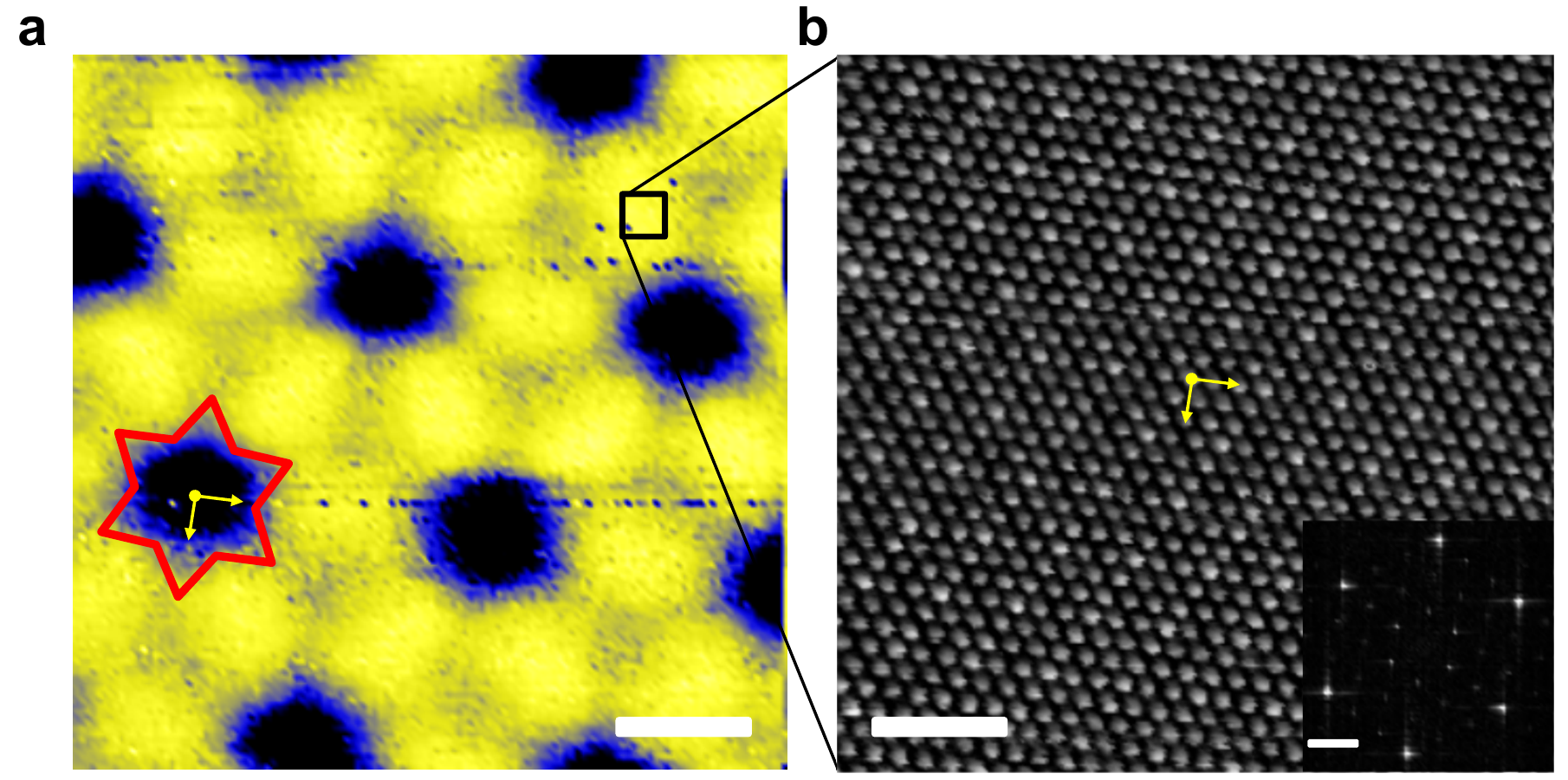}
\caption{In (\textbf{a}) we show an image of the vortex lattice taken at $T = 0.15$~K and magnetic field $B = 0.6$~T, parallel to the $c$ axis, obtained from the zero-bias tunneling conductance. The sixfold star shape of the single vortex is schematically indicated by a red star. White bar is 30 nm long. In (\textbf{b}) we show an atomic scale topographic STM image taken in the area highlighted by a black square in (\textbf{a}). White bar is 1.8 nm long. We indicate a lattice direction ($a$ axis of the hexagonal structure) and a direction in between crystalline axes by yellow arrows. The lower right inset shows the Fourier transform of the image (white bar is 1.9 nm$^{-1}$ long).}
\label{fs3}
\end{figure}

In Fig.~\ref{fs3} we show vortex core images at zero bias with the magnetic field parallel to the $c$ axis and perpendicular to the surface (Fig.~\ref{fs3}a). The surface Se atomic lattice image, typically observed in the topography of this compound is shown in Fig.~\ref{fs3}b. The vortex lattice is aligned with the atomic lattice.  We plot two vectors (yellow arrows in Fig.~\ref{fs3}), one is oriented parallel to an in-plane crystal axis, which corresponds to an azimuthal angle of $\varphi = 0^{\circ}$ and also parallel to one of the main axis of the vortex lattice. The other vector lies in between crystal axis vectors ($\varphi = 30^{\circ}$), parallel to the rays of the star observed in vortex lattice images at zero bias (see Fig.~\ref{fs2}a).

\section*{The tilted vortex core}

\begin{figure}[b]
\centering
\includegraphics[width=0.8\columnwidth]{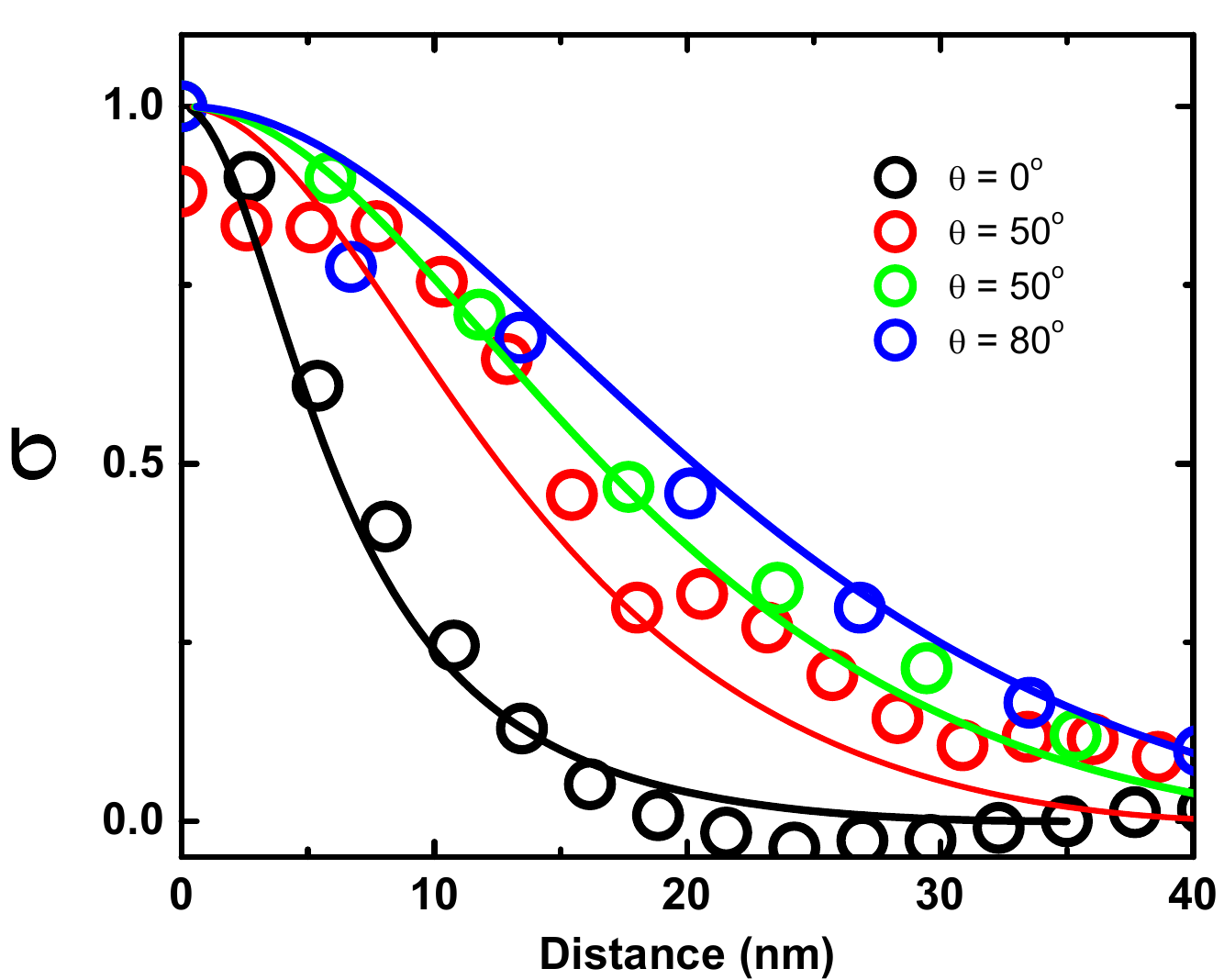}
\caption{Normalized tunneling conductance $\sigma$ radially averaged for several vortex cores for different polar tilt angles. Data are obtained from Fig.~\ref{f1}a of the main text and lines are fits according to Ref.~\cite{s_Fente16} as described in the text. The distance is the measured distance from the vortex center multiplied by the vortex frame intervortex spacing $a_0/\gamma$, where $\gamma$ is the anisotropy parameter and $a_0$ the intervortex distance.
}
\label{fs4}
\end{figure}

We can plot vortex core profiles by making the radial average of the smoothed tunneling conductance $\sigma$ around isolated vortices. We can fit this to the modified inverse bell shaped expression given in Ref.~\cite{s_Fente16}. In Fig.~\ref{fs4}, we show the result for the particular experiment of Fig.~\ref{f1}a. Generally, the vortex core size increases by a factor between two and four for the zero bias conductance maps, depending on the azimuthal angle.

\begin{figure}[t]
\centering
\includegraphics[width=\columnwidth]{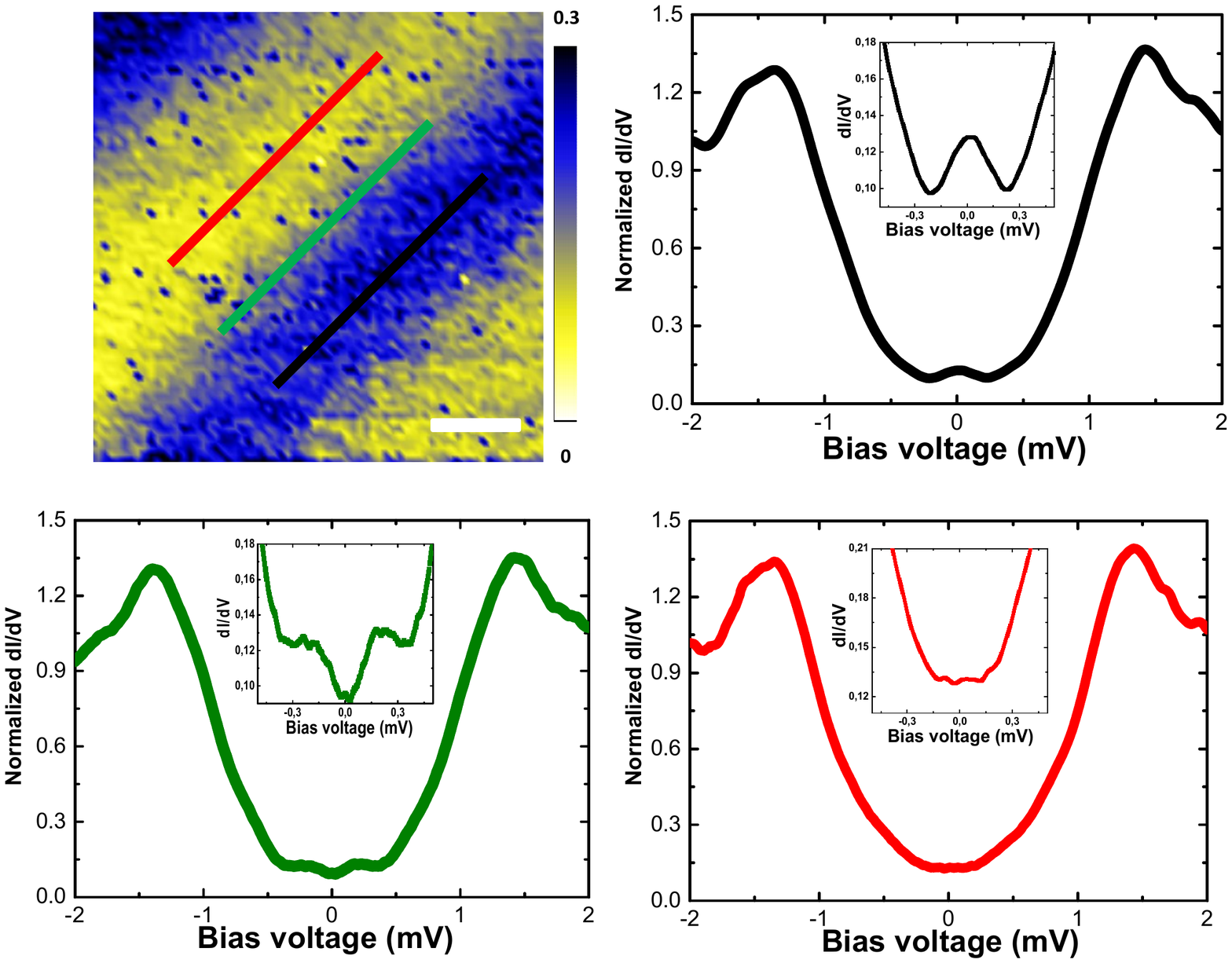}
\caption{We show a zoom into the stripes of Fig.~\ref{f1} of the main text. We make spatial averages of the tunneling conductance along the lines shown in the figure. We find a gap in-between stripes (red), a zero-bias peak at the centre of the stripes (black), which splits at the border of the stripes (green).}
\label{fs5}
\end{figure}

When making a zoom into the stripes discussed in Fig.~\ref{f1}b of the main text, we can gather tunneling conductance curves at the centre of the stripes, in-between stripes and on their sides. Such a zoom, with the corresponding tunneling conductance curves, is shown in Fig.~\ref{fs5}. We find a zero-bias peak at the centre of the stripes, which splits when leaving the stripes.

\begin{figure}[b]
\centering
\includegraphics[width=\columnwidth]{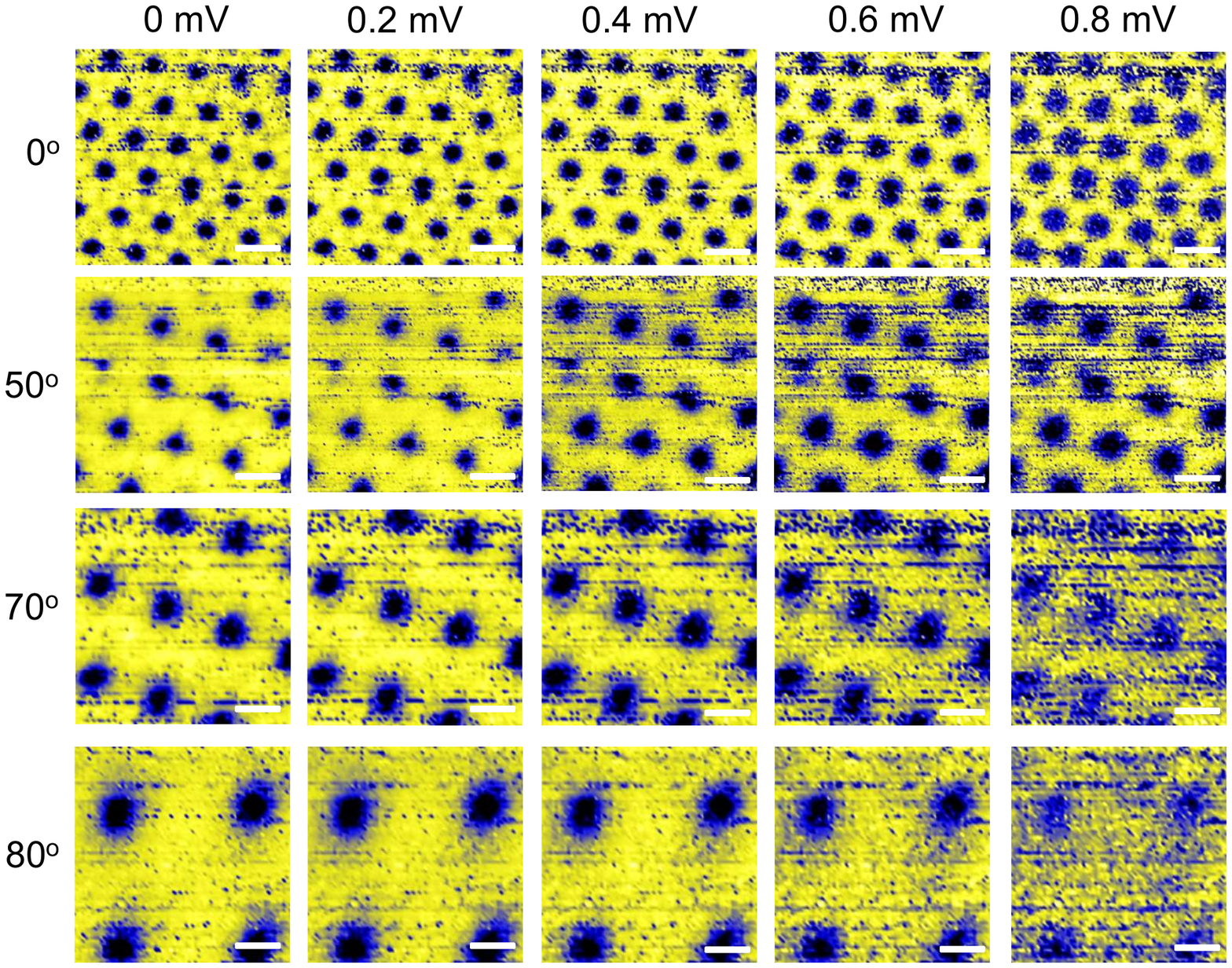}
\caption{Conductance maps of the vortex lattice as a function of the bias voltage (shown in the upper row) for different polar angles (shown in the left column). Scale bars are of 55 nm in all images and the color code is the same as in Fig.~\ref{f1} of the main text or Fig.~\ref{fs5} here.}
\label{fs6}
\end{figure}

In the figure Fig.~\ref{fs6} we show the bias voltage dependence of the conductance maps for different polar angles. When the field is parallel to the $c$ axis, we recover the well-known behavior of 2H-NbSe$_2$ and its characteristic changes in the vortex core patterns with bias voltage. We highlight here in particular that the apparent core size increases when reaching gap edge in the electronic DOS at about 0.8 mV. This is due to the superfluid currents around the vortex core that produce a Doppler shift in the quasiparticle peak in the DOS \cite{s_Kohen06,s_Maldonado13,s_Berthod13,s_Herrera17}. At high polar angles, the vortex cores are already very large at zero bias, indicating that the current distribution at the surface is highly non-trivial. This changes the core shape and its bias dependence, as discussed in the main text.

Furthermore, it is interesting to note that magnetic field nearly parallel to the hexagonal plane oriented along an in-plane direction which is neither along a crystal lattice axis nor in-between produces rather complex stripe patterns of the vortex lattice on the surface. In Fig.~\ref{fs7} we show a zero-bias tunneling conductance image obtained by a field tilted by $\varphi=15^{\circ}$ from the crystalline axis. The overall orientation of the stripes is along the tilt of the magnetic field, but the pattern has no clear periodicity, with stripes going along directions slightly tilted with the field and features that can be associated to either of the cases discussed in the main text.

\section*{Structure of the tilted vortex lattice}

\begin{figure}[tb]
\centering
\includegraphics[width=0.5\columnwidth]{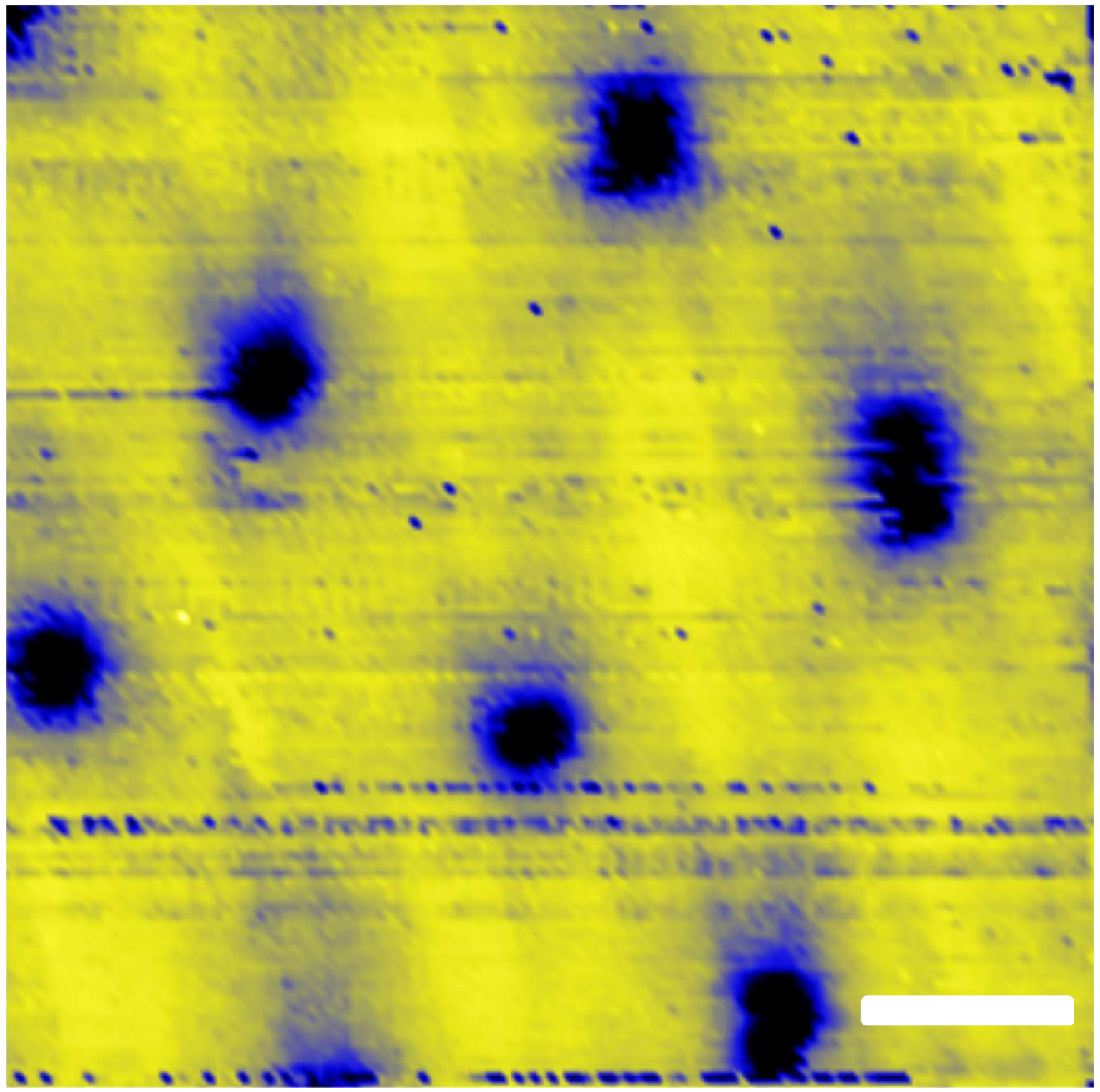}
\caption{Zero-bias conductance image with a magnetic field of 0.6~T nearly parallel to the surface and an azimuthal angle at $\varphi=15^{\circ}$ from a crystalline axis. We observe the same vortex and stripe pattern as discussed previously, although there are stripes going along oblique directions. Scale bar is of 120 nm.}
\label{fs7}
\end{figure}

Due to the out-of-plane anisotropy of 2H-NbSe$_2$ \cite{s_Foner73,s_Nader14,s_Xi15,s_Janssen98}, tilted magnetic fields produce considerable modifications in the vortex lattice. In particular, the lattice is no longer hexagonal but ellipsoidal when the magnetic field is tilted. The distortion is explained within Ginzburg-Landau (GL) theory of anisotropic superconductors \cite{s_Campbell88} and has been studied previously using STM and neutron scattering experiments \cite{s_Hess92,s_Hess94,s_Gammel94}. The vortex lattice is an elongated hexagon, with vortices located on an ellipse with its long axis of size $a_0/ \gamma$, with $\gamma=[1+\Gamma^{-2}\tan^2(\theta)]^{1/4}\cos^{1/2}(\theta)$ (with $a_0$ the intervortex distance, $\Gamma=\xi_{ab}/\xi_{c}$ the ratio of in-plane and out-of-plane coherence lengths and $\theta$ the polar tilt angle) \cite{s_Hess92,s_Hess94}. The hexagon within the ellipse has two possible orientations. The vertices of the hexagon may coincide either with the long axis or with the short axis of the ellipse. When the field is tilted by varying the polar angle ($\theta$) at an azimuthal angle ($\varphi$) fixed along one of the main vortex-lattice axis, at low tilts, the vertices coincide with the short axis of the ellipse. For high polar tilting angles, in particular for fields above $\theta=70^\circ$, a transition occurs in the vortex lattice frame. The vertices of the hexagon are no longer located along the short axis of the ellipse, because this situation is energetically unfavorable for high tilts \cite{s_Campbell88}. The vortex lattice then orients itself with the vertices of the hexagon coinciding with the long axis of the ellipse. This gives a rotation between A and B lattices when tilting the magnetic field and was found using STM and neutron scattering in different compounds with a strong out-of-plane anisotropy \cite{s_Hess92,s_Hess94,s_Gammel94,s_Kogan95,s_Fridman11,s_Fridman13}.

Note also that, for polar tilt angles $\theta$ close to 90$^{\circ}$, the intervortex distance at the surface frame diverges. This implies that minute changes in the orientation of the surface with respect to the magnetic field produce huge changes in the vortex lattice density at the surface. We show, in particular, areas of the same size at nominally the same $\theta$, with, five, eight and six vortices in the two panels of Fig.~\ref{f1}b and in Fig.~\ref{fs7}. Note in addition that these areas are quite large, close to the maximal scanning range of cryogenic STM experiments. The difference between numbers of vortices seen in each image might change when using larger scanning ranges, if surfaces with slightly different tilts can be obtained within the same field of view.

\end{document}